\documentstyle[12pt]{article}

\def\pasq{q} 
\def\Pasq{Q} 
\def\pas{h}  
\def\Pas{H}  
\def \Plv{Pain\-le\-v\'e}
\def \D {\hbox{d}}
\def \Log {\mathop{\rm Log}\nolimits}

\begin{document}

\pagestyle{plain} 


\begin{center}
 {\bf A NEW METHOD TO TEST DISCRETE PAINLEV\'E EQUATIONS}
\end{center}

\vskip 0.5 truecm

{\bf Robert Conte\dag} and {\bf Micheline Musette\ddag}

\medskip

\dag
Service de physique de l'\'etat condens\'e,
CEA Saclay,
\hfill \break \indent
F-91191 Gif-sur-Yvette Cedex,
France
\medskip

\ddag
Dienst Theoretische Natuurkunde,
Vrije Universiteit Brussel,
\hfill \break \indent
B-1050 Brussel,
Belgique

\vskip 0.5 truecm

\noindent {\it Short title} :  Testing discrete Painlev\'e equations
 
\vglue 1.0truecm
\baselineskip=12truept
 

\vskip 0.8 truecm

\noindent {\it Keywords}
\par discrete Painlev\'e property
\par discrete Painlev\'e test
\par discretization rules
 
\vskip 0.8 truecm

{\it {\bf Abstract} -- 
Necessary discretization rules to preserve the Painlev\'e property are
stated.
A new method is added to the discrete Painlev\'e test,
which perturbs the continuous limit
and generates infinitely many no-log conditions.
}

\vfill 
\noindent 

\rightline{\noindent Phys.~Lett.~A 
           \hskip 1 truemm 31 May 1996, revised 7 October 1996
           \hskip 5 truemm S96/032 (solv-int/9610007)}

\eject


\baselineskip=14truept 


\section{Introduction}
\indent

The point of view developed in this article is,
given some continuous differential equation,
to find its discretizations which preserve some global property,
namely the explicit linearizability or more generally
the {\it discrete Painlev\'e property},
a notion defined section \ref{sectiondPP}
together with its group of invariance.
From this point of view,
we will never have to address the question of finding a continuous limit,
and consequently we will always write the discrete equations
in a form as close as possible to the canonical form of the continuous limit,
which is well established~\cite{PaiActa,GambierThese}.

Discrete equations can be considered as
functional equations linking the values taken by some
field variable $u$ at a finite number $N+1$ of points,
either arithmetically consecutive~: $x + k \pas, k-k_0=0,1, \dots, N$,
or      geometrically consecutive~: $x \pasq^k,  k-k_0=0,1, \dots, N$,
where $\pas$ or $\pasq$ is the lattice stepsize,
assumed to lay in some neighborhood of, respectively, $0$ or $1$,
and $k_0$ is just some convenient origin.
The integer $N$ is called the {\it order} of the equation.
They have received a decisive impulse from statistical physics in 1990
with the discovery of a discrete analog of the first Painlev\'e equation (P1)
\cite{BrezinKazakov,DouglasShenker,GrossMigdal}
\begin{eqnarray}
& &
E \equiv
-(\overline{u} - 2 u + \underline{u})/ \pas^2
+ 2 (\overline{u} + u + \underline{u}) u + x =0
\label{eqdP1BrezinKazakov}
\end{eqnarray}
and of the second Painlev\'e equation (P2)~\cite{PS1990,NP1991}
(in the particular case $\alpha=0$)
\begin{eqnarray}
& &
E \equiv
-(\overline{u} - 2 u + \underline{u}) / \pas^2
+ (\overline{u} + \underline{u}) u^2 + x u + \alpha=0.
\label{eqdP2PeriwalShevitz}
\end{eqnarray}
The above short notation
\begin{eqnarray}
& &
                      u  = u(x      ),\
            \overline{u} = u(x +   h),\
           \underline{u} = u(x -   h),\
 \overline{ \overline{u}}= u(x + 2 h),\
\dots
\hskip 7 truemm
\end{eqnarray}
is adopted throughout the article.
These two equations clearly admit as
continuous limits ($\pas \to 0$) (P1) and (P2)
\begin{eqnarray}
& &
E \equiv
-\D^2 u / \D x^2 + 6 u^2 + x = 0,
\label{eqP1}
\\
& &
E \equiv
-\D^2 u / \D x^2 + 2 u^3 + x u + \alpha= 0.
\label{eqP2}
\end{eqnarray}
In fact the equation (\ref{eqdP1BrezinKazakov}) had been written down
by many authors
\cite{Shohat,HarishChandra,Freud,Bessis1979,IZ1980},
but ref.~\cite{BrezinKazakov,DouglasShenker,GrossMigdal}
were the first to notice the continuous limit to (P1).

The reason why these two discrete equations deserve the name of
discrete Painlev\'e equations, in short d--(P1) and d--(P2),
is that they possess the discrete Painlev\'e property.
Indeed, both admit a discrete Lax pair
$(A,B,z,\psi,\pas)$~\cite{IKF1990}
\begin{eqnarray}
& &
\psi(x + \pas/2) = A \psi(x - \pas/2),\
\partial_z \psi(x - \pas/2) = B \psi(x - \pas/2).
\label{eqLaxDiscretCommutateur}
\end{eqnarray}
The matricial Lax pair of
the d--(P2) (\ref{eqdP2PeriwalShevitz}) is~\cite{JBH1992,GNPRS1994}
{\small
\begin{eqnarray}
& &
A= \pmatrix {1/z & \pas u \cr \pas u & z \cr},\
t={z - z^{-1} \over 2 \pas},\
z=1 + \pas t + O(\pas^2)
\nonumber
\\
& &
\pas z B=
(-2 u \underline{u} + 4 t^2 - (x - \pas /2)) \pmatrix{1 & 0 \cr 0 & 1 \cr}
+ (z+1/z) (u-\underline{u})/ \pas \pmatrix{0 & 1 \cr -1 & 0 \cr}
\nonumber
\\
& &
\phantom{\pas z B=}
- (2 t (u+\underline{u}) + \alpha/t) \pmatrix{0 & 1 \cr 1 & 0 \cr}
\label{eqLaxdP2PeriwalShevitz}
\\
& &
\partial_z A + A B - \overline{B} A 
 = 2 z \pmatrix{0 & 1 \cr -1 & 0 \cr} E,
\nonumber
\end{eqnarray}
}
while the matricial form of that of 
the d--(P1) (\ref{eqdP1BrezinKazakov})
is
\begin{eqnarray}
& &
A=\pmatrix{1-\pas^2 z & \pas (2 u - 2 t) \cr \pas & 1-\pas^2 z \cr},\
t=z - \pas^2 z^2/2,\
z=t + (t^2/2) \pas^2 + O(\pas^4)
\nonumber
\\
& &
B= \pmatrix{
  2 (1 - \pas^2 z) (u - \underline{u}) / \pas &
-4 u \underline{u} -2(x - \pas /2) + 4 t (u + \underline{u}) - 16 t^2 \cr
2 (u + \underline{u}) + 8 t &
 -2 (1 - \pas^2 z) (u - \underline{u}) / \pas \cr},
\label{eqLaxdP1BrezinKazakov}
\\
& &
\pas^{-1} (\partial_z A + A B - \overline{B} A)
 = 2 \pmatrix{1 & 0 \cr 0 & -1 \cr} E,
\nonumber
\end{eqnarray}
a new result to our knowledge.


Just like its continuous counterpart,
the discrete Painlev\'e test is the set of all methods one can imagine
to build necessary conditions for a given discrete equation to possess
the discrete Painlev\'e property.
At the present time, only one such method is known,
namely the singularity confinement method of Grammaticos {\it et al.}
\cite{GRP1991}.
We propose here a new method,
based on the perturbation theorem of Poincar\'e,
which is applicable to differential systems
in an arbitrary number of independent variables,
whether discrete, continuous or even mixed discrete--continuous.
As an example,
we apply it to a qualitative discretization of (P1) and isolate three
candidates d--(P1).
\smallskip

The paper is organized as follows.
The discrete Painlev\'e property (PP) and its group of invariance
are defined in section \ref{sectiondPP},
and a first set of basic rules of discretization is given.
Section \ref{sectionTestDiscret}
explains the new method for the discrete Painlev\'e test,
introduces a new notion,
the {\it depth} of a discrete equation,
and applies the method to qualitative candidates d--(P1) and d--(P2).

\section{The discrete Painlev\'e property, its group of invariance,
basic rules of discretization}
\label{sectiondPP}
\indent

The (continuous) Painlev\'e property is defined~\cite{PaiLecons}
as the absence of movable critical singularities in the general solution of
a differential equation
\begin{eqnarray}
& &
\forall x\ :\
E(x,u,\D u / \D x,\dots,\D^N u / \D x^N)=0,
\label{eqContinuous}
\end{eqnarray}
where a singularity is said {\it movable} (as opposed to {\it fixed})
if its location in the complex plane of $x$ depends on the initial conditions,
and {\it critical} if the solution is multivalued around it.
For shortness, following Bureau~\cite{Bureau1939},
we will use the terms ``stability'' for PP, 
``stable'' or ``unstable'' for an equation with or without the PP.

The PP is invariant under the group of birational transformations
\begin{eqnarray}
& &
(u,x) \to (U,X) :\
u=r(x,U,\D U / \D X,\dots,\D^{N-1} U / \D X^{N-1})=0,\
x=\Xi(X),
\\
& &
(U,X) \to (u,x) :\
U=R(X,u,\D u / \D x,\dots,\D^{N-1} u / \D x^{N-1})=0,\
X=\xi(x),\
\label{eqGroupBirationalContinuous}
\end{eqnarray}
($r$ and $R$ rational in $U,u$ and their derivatives,
analytic in $x,X$).
An easier to manage subgroup is made of the homographic transformations
\begin{eqnarray}
& &
(u,x) \to (U,X) :\
u={a U + b \over c U + d},\
X=\xi(x),\
a d - b c \not=0
\label{eqGroupHomographicContinuous}
\end{eqnarray}
where $(a,b,c,d,\xi)$ are arbitrary analytic functions of $x$.
In his classification of second order first degree equations,
Gambier~\cite{GambierThese} has found respectively twenty-four and fifty
 equivalence classes for these two groups 
(with minor later corrections~\cite{BureauMI,CosScou}).

In the discrete case, 
let us consider equations 
\begin{eqnarray}
& &
\forall x\ \forall \pas\ :\
E(x,\pas,\{u(x+k \pas),\ k-k_0=0,\dots,N\})=0
\label{eqDiscretexpas}
\\
& &
\forall x\ \forall \pasq\ :\
E(x,\pasq,\{u(x \pasq^k),\ k-k_0=0,\dots,N\})=0
\label{eqDiscretexpasq}
\end{eqnarray}
algebraic in the values of the field variable,
with coefficients analytic in $x$ and the stepsize.
It should be noted that $u$ is a function of
{\it two} variables, $x$ and the stepsize.
A natural definition for the discrete Painlev\'e property,
which seems to have never been given before, is the following.

{\it Definition}.
A discrete equation is said to possess the
{\it discrete Painlev\'e property}
if and only if
there exists a
neighborhood of $\pas=0$ (resp.~$\pasq=1$)
at every point of which
the general solution $x \to u(x,\pas)$ (resp.~$x \to u(x,\pasq)$)
has no movable critical singularities.

{\it Remarks}.
\begin{enumerate}
\item
The definition reduces to that of the continuous PP in the continuous limit.

\item
This definition immediately extends to equations in an arbitrary number of
independent variables, discrete or continuous,
the extension starting then from the definition of the PP suited to 
partial differential equations (PDEs),
which we do not remind here since this is not our subject.

\end{enumerate}

The discrete PP is invariant under the discrete analog of
(\ref{eqGroupBirationalContinuous}),
which is the group of nonlocal discrete birational transformations
\begin{eqnarray}
& &
u=r(x,\pas \hbox{ or }\pasq,U,\overline{U},\underline{U},\dots),\
\nonumber
\\
& &
U=R(X,\Pas \hbox{ or }\Pasq,u,\overline{u},\underline{u},\dots),\
X=\xi(x,\pas \hbox{ or }\pasq),\
\Pas =\eta(\pas),\
\Pasq=\kappa(\pasq),\
\label{eqGroupBirationalDiscrete}
\end{eqnarray}
($r$ and $R$ rational in $U,\overline{U},\underline{U},\dots,
u,\overline{u},\underline{u},\dots$,
analytic in $x$ and the stepsize,
$\xi,\eta,\kappa$ analytic).
There exist two discrete analogs of the subgroup
(\ref{eqGroupHomographicContinuous}),
and both may be useful to establish the discrete equivalent of the
classification of Gambier.
The first one is the group of nonlocal transformations
(\ref{eqGroupBirationalDiscrete})
which in the continuous limit reduce to the homographic transformations
(\ref{eqGroupHomographicContinuous}),
where $(a,b,c,d,\xi)$ are arbitrary analytic functions of $x$ and of the
stepsize.
The second one is the group of local homographic transformations
($r$ and $R$ homographic in $U$ and $u$,
independent of 
$\overline{U},\underline{U},\dots,\overline{u},\underline{u},\dots$,
analytic in $x$ and the stepsize,
$\xi,\eta,\kappa$ analytic).

{\it Remark}.
The first subgroup seems more useful,
although it does not contain the transformation
$u=\pas^k U,\ k \in {\cal Z}$.

Let us now give some basic rules for discretizing a given continuous
equation (\ref{eqContinuous})
into either a difference equation (\ref{eqDiscretexpas})
or a $q$--difference equation (\ref{eqDiscretexpasq}).

The question of discretization is well known in numerical analysis,
where one looks for a {\it scheme} of discretization
which maximizes the order, called {\it scheme order},
of the remainder of the expansion of the left-hand side of
(\ref{eqDiscretexpas})
in a Taylor series of $\pas$ around the center of the $N+1$ points.
A scheme of discretization is said {\it exact} iff it has an infinite order,
like the one for the linear ODE $u^{(N)}=0$
\begin{equation}
\sum_{j=0}^{N} (-1)^j C_N^j u(x + j \pas)=0.
\end{equation}

Contrary to numerical analysis,
only interested in a {\it local} integration,
we require the scheme of discretization to
preserve the differential order $N$,
an essential element for a {\it global} knowledge of the solution :
every discretization must involve $N+1$ points.

Just like in the continuous case,
there exists another important element concerning discretization rules.

{\it Definition}.
The {\it degree} of a discrete equation is the highest of the two
polynomial degrees of the LHS $E$ of the equation in
$u(x)$ and $u(x + N \pas)$, or $u(x)$ and $u(x {\pasq}^N)$,
where $E$ is assumed polynomial in the $N+1$ variables 
$u(x + k \pas)$ or $u(x q ^k)$.

Let us have a closer look at the degree before and after a discretization.

Consider first a second order, first degree equation 
with a single valued general solution
\begin{eqnarray}
& &
u=(c_1 x + c_2)^2,\
u u'' - (1/2) u'^2 =0,
\end{eqnarray}
and its exact discretization
\begin{eqnarray}
& &
u {\overline{u} - 2 u + \underline{u} \over \pas^2}
-(1/2) {(\overline{u} - u) (u - \underline{u}) \over \pas^2}
-(1/8) \pas^2
\left({\overline{u} - 2 u + \underline{u} \over \pas^2}\right)^2=0.
\label{eqExempleExact}
\end{eqnarray}
Although the exact discretization does not conserve the degree,
there exists a second order scheme which conserves it : 
this is the one defined by dropping the third term in (\ref{eqExempleExact})
\begin{eqnarray}
& &
u {\overline{u} - 2 u + \underline{u} \over \pas^2}
-(1/2) {(\overline{u} - u) (u - \underline{u}) \over \pas^2}
=0.
\label{eqExempleTronque}
\end{eqnarray}
What is remarkable is that both schemes are explicitly linearizable into
$\overline{\psi} - 2 \psi + \underline{\psi}=0$
by the respective transformations
\begin{eqnarray}
& &
u=\psi^2,\
\\
& &
u=(1/4)
[\psi(\overline{\psi}+\underline{\psi})
 +\psi^2 +\overline{\psi} \underline{\psi}].
\end{eqnarray}

Consider next a first order, first degree ODE with a multivalued general
solution
\begin{equation}
u=(x-x_0)^{-1/2},\
u' + (1/2) u^3 = 0,
\end{equation}
and its exact discretization
\begin{equation}
{\overline{u} - \underline{u} \over \pas} + 
{\overline{u}^2 \underline{u}^2 \over \overline{u} + \underline{u}} = 0.
\end{equation}
There exists no discretization scheme conserving the degree one of the
continuous equation,
and the reason is here the multivaluedness.

We therefore conjecture : 
given an algebraic differential equation with the PP,
there exists a discretization scheme of order two which conserves the degree.

Hence the additional rule,
restricted to algebraic differential equations with the PP : 
every second order scheme must also conserve the degree.


\section{The discrete Painlev\'e test
\label{sectionTestDiscret}}
\indent

In the continuous case,
{\it all} the methods of the Painlev\'e test,
without exception,
are based on two theorems and only two,
namely
the existence theorem of Cauchy
and the theorem of perturbations of Poincar\'e~\cite{Poincare}.
This is explained in detail in
the lecture notes of a Chamonix school~\cite{Chamonix1993}.

The discrete case is an arithmetics problem,
hence very difficult.
But a usual approach in arithmetics is to convert the problem
into one of analysis,
for which all the tools dealing with the notion of limit are available.
This is appropriate here,
in order to make applicable the perturbation theorem of Poincar\'e.

Consider an arbitrary discrete equation (\ref{eqDiscretexpas}),
also depending on some parameters $a$,
and 
let $(x, \pas, u, a) \to (X, \Pas, U, A, \varepsilon)$
be an arbitrary perturbation admissible by the theorem of Poincar\'e
(which excludes any nonanalyticity,
like $\varepsilon^{1/5}$, for the perturbed variables).
A necessary condition is that the limit $\varepsilon=0$ possesses the PP
(discrete or continuous, this does not matter).

In order to illustrate the method, 
let us discretize the two equations (P1) and (P2)
by a second order scheme,
using the rules previously stated.
We will detail the case of (P1) and just give the result for (P2).

The candidates d--(P1) and d--(P2) are as follows
(we choose them rather simplified to alleviate the presentation)
\begin{eqnarray}
E & \equiv & 
- (\overline{u} - 2 u + \underline{u}) \pas^{-2} 
+ 3 \lambda_1 (\overline{u} + \underline{u}) u
+ 6 \lambda_2 u^2
+ 6 \lambda_3 \overline{u} \underline{u}
+ g=0.
\label{eqP1CompleteDiscrete}
\\
E & \equiv & 
- (\overline{u} - 2 u + \underline{u}) \pas^{-2} 
+ \lambda_1 (\overline{u} + \underline{u}) u^2
+ 2 \lambda_2 u^3
+ 2 \lambda_3 \overline{u} \underline{u} u
\nonumber
\\
& &
+ x (\mu_1 (\overline{u} + \underline{u})/2 + \mu_2 u)
+ \alpha=0,
\label{eqP2CompleteDiscrete}
\end{eqnarray}
with $\sum \lambda_k=1,\sum \mu_k=1$,
$\alpha$ a constant and $g$ an unspecified function of $x$.

The test will generate necessary conditions on
$(\lambda_i, \mu_i, g)$.
One must find at least the following solutions,
where the equations have a Lax pair :
for d--(P1),
$g=x$ and 
$\overrightarrow \lambda=(2/3,1/3,0)$;
for d--(P2),
$\overrightarrow \lambda=(1,0,0), \overrightarrow \mu=(0,1)$.
One may also find the following solutions,
isolated by the singularity confinement method
although no Lax pair is yet known :
for d--(P1),
$g=x$ and 
$\overrightarrow \lambda=(1,0,0)$ (ref.~\cite{GRP1991}),
$\overrightarrow \lambda=(1/2,1/4,1/4)$ (ref.\cite{RG96} eq.~(5.5)).

The method which we propose is
a perturbation of the continuous limit entirely analogous to
the Fuchsian~\cite{CFP1993} or nonFuchsian~\cite{MC1995} 
perturbative method of the continuous case.
This perturbation is defined by an expansion of $u$ 
as a Taylor series in $\varepsilon$,
where $\varepsilon$ is nothing else than the lattice stepsize
\begin{eqnarray}
& &
x      \hbox{ unchanged},\
\pas = \varepsilon,\
\pasq = e^{\varepsilon},\
u    = \sum_{n=0}^{+ \infty} \varepsilon^n u^{(n)},\
a    = \hbox{ analytic }(A, \varepsilon).
\label{eqPerturbationLimiteContinue}
\end{eqnarray}
It generates an infinite sequence of (continuous) differential equations
$E^{(n)}=0$
defined by
\begin{eqnarray}
& &
E = \sum_{n=0}^{+ \infty} \varepsilon^n E^{(n)}
\\
& &
E^{(n)} (x, u^{(0)}, \dots, u^{(n)})
\equiv
 {E^{(0)}(x, u^{(0)})}' u^{(n)} + R^{(n)}(x, u^{(0)}, \dots, u^{(n-1)}) = 0,\ 
n\ge 1,
\nonumber
\end{eqnarray}
whose first one $n=0$ is the ``continuous limit''.
The next ones $n\ge 1$, which are linear inhomogeneous,
have the same homogeneous part $E^{(0)'} u^{(n)}=0$ independent of $n$,
defined by the derivative of the equation of the continuous limit,
while their inhomogeneous part $R^{(n)}$ (``right-hand side'')
comes at the same time from the nonlinearities and the discretization.

The only tiny difference with the continuous perturbative method
lies in the explicit dependence on $\varepsilon$
of the unperturbed equation.
This does not affect the rest of the method and,
depending on the nature, Fuchsian or nonFuchsian,
of the linearized equation $E^{(1)}=0$ at a singulier point of $u^{(0)}$,
one then applies, without any other change,
either the Fuchsian perturbative method of the continuous case
(Ref.~\cite{CFP1993} p.~42)
or  the nonFuchsian perturbative method of the continuous case
(Ref.~\cite{MC1995} p.~342).
We adopt the notation used in these two articles.

As a tutorial introduction,
let us first handle the Euler scheme for the Bernoulli equation 
\begin{equation}
E \equiv (\overline{u} - u) / \pas + u^2 = 0
\label{eqdRiccatiEuler}
\end{equation}
i.e.~the logistic map of Verhulst,
a paradigm of chaotic behaviour.
Let us expand the terms of (\ref{eqdRiccatiEuler})
according to the perturbation (\ref{eqPerturbationLimiteContinue})
up to an order in $\varepsilon$ sufficient to build the first 
equation $E^{(1)}=0$ beyond the continuous limit $E^{(0)}=0$
\begin{eqnarray}
u & = &
u^{(0)} + u^{(1)} \varepsilon + O(\varepsilon^2)
\\
u^2 & = &
u^{(0)^2}
 + 2 u^{(0)} u^{(1)} \varepsilon 
 + O(\varepsilon^2)
\\
\overline{u} & = &
u + u' \pas + (1/2) u'' \pas^2 + O(\pas^3)
\\
{\overline{u} - u \over \pas}
& = &
u^{(0)'} + (u^{(1)'} + (1/2) u^{(0)''}) \varepsilon + O(\varepsilon^2).
\end{eqnarray}
The equations of orders $n=0$ et $n=1$ are written as
\begin{eqnarray}
E^{(0)} & = &
u^{(0)'} + u^{(0)^2} =0
\\
E^{(1)} & = &
E^{(0)'} u^{(1)} + (1/2) u^{(0)''} = 0,\
E^{(0)'} =\partial_x + 2 u^{(0)}.
\end{eqnarray}
Their general solution is
\begin{eqnarray}
u^{(0)} & = &
\chi^{-1}, \chi=x-x_0,\ x_0 \hbox{ arbitrary}
\\
u^{(1)} & = &
u_{-1}^{(1)} \chi^{-2} - \chi^{-2} \Log \psi,\ 
\psi=x-x_0,\
u_{-1}^{(1)} \hbox{ arbitrary},
\end{eqnarray}
and the movable logarithm proves the instability as soon as order $n=1$,
at the Fuchs index $i=-1$.

{\it Definition}.
One calls {\it depth} of a discrete equation
the rank $n$ of the first Taylor coefficient of $u$ in series of the stepsize
not to possess the continuous PP.

The highest the depth,
the largest the choice of the stepsize for a numerical integration.
Equal to one only for the Verhulst map,
the depth is infinite for any stable equation.

{\it Remarks}.
\begin{enumerate}

\item
The depth is bounded from above by the smallest of the perturbative orders $n$
at which a movable logarithm appears
when one runs over the several families of movable singularities.

\item
The only restriction on $u^{(0)}$ is not to be what is called
a singular solution
(not obtainable from the general solution by assigning values to the
arbitrary data),
i.e.~it can be either the general solution (as above) or a particular one,
it can also be either global (as above) or local (Laurent series).

\item
The derivative of  $E^{(0)}$ at $u^{(0)}$ having $N$ as differential order
since the singular solutions are excluded,
the sum $u^{(0)} + u^{(1)} \varepsilon$ is a local representation
of the {\it general} solution in the neighborhood of 
$(\chi,\varepsilon) = (0,0)$.
One must of course at the order $n=1$ include in $u^{(1)}$ all the
arbitrary data missing in $u^{(0)}$,
and only them.
Here, no such arbitrary need be included,
and one can set to zero the term $u_{-1}^{(1)} \chi^{-2}$,
i.e.~the partial derivative of $u^{(0)}$ with respect to the arbitrary data
$x_0$.

\end{enumerate}

For the example (\ref{eqP1CompleteDiscrete}), 
one obtains
\begin{eqnarray}
E^{(0)} & = &
-{u^{(0)}}'' + 6 u^{(0)^2} + g(x)=0
\label{eqP1discretOrder0}
\\
E^{(0)'} & = & - \partial_x^2 + 12 u^{(0)},\
\\
R^{(1)} & = & 0
\\
R^{(2)} & = &
6 u^{(1)^2} - 6 \lambda_3 u^{(0)'^2}
+ 3 (\lambda_1 + 2 \lambda_3) u^{(0)} u^{(0)''}
-(1/12) {u^{(0)''''}},
\end{eqnarray}
etc.
At order $n=0$ (the ``continuous limit''),
Painlev\'e has found the necessary condition $g''=0$.
Let us choose $g(x)=x + b$ and for $u^{(0)}$ the local general solution of
(\ref{eqP1discretOrder0})
\begin{eqnarray}
u^{(0)} & = &
\chi^{-2} (1 - (b/10) \chi^4 - (1/6) \chi^5 + u^{(0)}_{6} \chi^6 + \dots)
\label{eqP1discretOrder0Expansion}
\end{eqnarray}
in which $x_0$ and $u^{(0)}_{6}$ are arbitrary.
This is then the Fuchsian perturbative method which must be used
\begin{eqnarray}
n \ge 1 : \
u^{(n)} & = &
\sum_{j=-n}^{+ \infty} u^{(n)}_{j} \chi^{j+p},\
u^{(n)}_{-1}=0,\
u^{(n)}_{6}=0,\
E^{(n)} = 
\sum_{j=-n}^{+ \infty} E^{(n)}_{j} \chi^{j+q},
\end{eqnarray}
in order to generate the necessary conditions 
\begin{eqnarray}
n \ge 1 : \
Q^{(n)}_{i} & \equiv & E^{(n)}_{i} =0,\
i \in \{-1,6\}.
\end{eqnarray}
The first nonzero conditions are
\begin{eqnarray}
Q^{(4)}_{6} \equiv
& &
3 (- \lambda_2 + 3 \lambda_2^2 + 10 \lambda_3 
       + 21 \lambda_2 \lambda_3 - 60 \lambda_3^2)/50 = 0
\\
Q^{(6)}_{-1} \equiv 
& &
9 (- \lambda_2 + 2 \lambda_2^2 + 3 \lambda_2^3 
     - 19 \lambda_3 + 28 \lambda_2 \lambda_3
\nonumber
\\
& &
 + 11 \lambda_2^2 \lambda_3 + 58 \lambda_3^2
 - 7 \lambda_2 \lambda_3^2 - 39 \lambda_3^3)/5 = 0
\\
Q^{(8)}_{6} \equiv 
& &
b f_4 
=0
\\
Q^{(10)}_{-1} \equiv 
& &
b f_5
=0
\\
Q^{(10)}_{6} \equiv 
& &
u^{(0)}_{6} f_5
=0
\\
Q^{(12)}_{-1} \equiv 
& &
u^{(0)}_{6} f_6
=0
\\
Q^{(12)}_{6} \equiv 
& &
b^2 f_6 
=0
\end{eqnarray}
in which the $f_k$'s, all different, represent polynomials
of degree $k$ in $(\lambda_1,\lambda_2,\lambda_3)$.

The polynomials $Q^{(n)}_{i}$ in the variables 
$(\lambda_1,\lambda_2,\lambda_3)$
are generated by the ideal
\begin{eqnarray}
& &
\lambda_1  + \lambda_2 + \lambda_3 -1,\
\lambda_3 (\lambda_2 - \lambda_3),\
\lambda_3 (\lambda_1 - 2 \lambda_2),\
\lambda_2 (\lambda_1 - 2 \lambda_2),
\end{eqnarray}
and this ideal possesses the only three zeroes :
\begin{eqnarray}
(\lambda_1,\lambda_2,\lambda_3)
& = &
(2/3,1/3,0),\
(1,0,0),\
(1/2,1/4,1/4).
\label{eqP1CompleteDiscreteNecessary}
\end{eqnarray}
This result is obtained at $(n,i)=(6,-1)$,
but one must proceed to $(n,i)=(8,6)$ to get rid of three
irrational extraneous solutions;
we have checked the vanishing of all no-log conditions up to
$(n,i)=(22,6)$.

The first value $\overrightarrow \lambda=(2/3,1/3,0)$
(case $a=1$ in Ref.~\cite{GRP1991})
corresponds to the d--(P1) (\ref{eqdP1BrezinKazakov}) with a Lax pair 
found in quantum gravity, the condition is then sufficient.

The second value $(1,0,0)$ 
(case $a=0$ in Ref.~\cite{GRP1991})
corresponds to a candidate d--(P1) which admits the second order matricial 
Lax pair
\begin{eqnarray}
& &
A=\pmatrix{a_1 & 2 \pas (f_2 u -f_3) \cr 0 & -a_1 \cr},\
\nonumber
\\
& &
B_{2,1}=0,\
B_{1,1}=-B_{2,2}=4 J (\pas^{-3} a_1 / f_2 + f_2' / (2 J f_2)),\
\nonumber
\\
& &
B_{1,2}=
J(- 12 u \underline{u} - 2(x - \pas /2) + 4 \pas^{-2} (u + \underline{u})
  - 8 \pas^{-2} f_3/f_2)
\nonumber
\\
& &
\phantom{B_{1,2}=}
 + \pas a_1^{-1} (f_3' - (f_3/f_2) f_2')
\nonumber
\\
& &
\partial_z A + A B - \overline{B} A 
 = \pmatrix{0 & -4 a_1 J \cr 0 & 0 \cr} E,
\label{eqLaxdP1bis}
\end{eqnarray}
in which $J,f_2,f_3$ are arbitrary functions of $z$,
and $a_1$ an arbitrary constant;
however,
we have not succeeded in finding a continuous limit to this Lax pair.

The third value $(1/2,1/4,1/4)$
defines an equation $E(u)=0$ equivalent to that $E(v)=0$ for $(1,0,0)$ 
under the discrete birational transformation~\cite{RG96} 
\begin{eqnarray}
& &
2 v(x)= u(x + \pas/2) + u(x - \pas/2)
\nonumber
\\
& &
2 u(x)=-3 \pas^2 v(x -\pas/2) v(x + \pas/2)
+ v(x -\pas/2) + v(x + \pas/2) - \pas^2 x/2.
\end{eqnarray}

As to the scheme (\ref{eqP2CompleteDiscrete}) for d--(P2),
it admits the unique solution
$\overrightarrow \lambda=(1,0,0)$,
$\overrightarrow \mu=(0,1)$,
obtained, 
for the family $u^{(0)} \sim \pm \chi^{-1}$ with Fuchs indices $(-1,4)$,
at $(n,i)=(4,4)$.
This is equ.~(\ref{eqdP2PeriwalShevitz}), so the condition is also sufficient.

{\it Remarks}.
\begin{enumerate}
\item
These four solutions are simple rational numbers,
and moreover they are convex : 
$\forall k :\ 0 \le \lambda_k \le 1,\ 0 \le \mu_k \le 1$.

\item
Following Painlev\'e~\cite{PaiBSMF},
one should in fact search for the ``complete equation'',
i.e.~for all the admissible nondominant terms which can be added to
the candidate d--(P1) (\ref{eqP1CompleteDiscrete}) without destroying the PP.
This will be done in future work.

\end{enumerate}

\subsection{Discussion}
\indent

The example (\ref{eqP1CompleteDiscrete}) has been chosen for its simplicity
but it already allows for some comparison
between the singularity confinement method 
and the perturbation of the continuous limit.

When applied to (\ref{eqP1CompleteDiscrete}),
the singularity confinement method finds the same necessary conditions 
(\ref{eqP1CompleteDiscreteNecessary}),
as we have checked
(this was checked earlier with different types of equations
\cite{GRP1991,RG96}).
So the most important point is,
again for this particular example,
the identity of the set of necessary conditions generated by both methods.
Other examples are currently under investigation
\cite{LabrunieConteChazyIII,LabrunieThese}
to detect situations where the two methods would produce complementary,
not identical results.

We have no explanation for the apparent identity of the results
produced by the two methods,
but a reason could be the existence of a perturbation of Poincar\'e
identical to the method of confinement,
and whose no-log conditions would be identical to the 
confinement conditions.
Despite a thorough search for such a perturbation, we have not found one.

Since it only relies on the existence of a continuous limit,
our method can be extended without difficulty
to equations in an arbitrary number of dependent or independent variables,
whether the equations be discrete or mixed continuous and discrete.

The singularity confinement method has also been extended to such situations
\cite{RGT92,RGT93};
however, in the case of $m$ discrete independent variables,
one must check in addition 
that the result of the iteration is independent of the path followed 
on the $m-$th dimensional lattice.

As pointed out by a referee,
the present method is also applicable to a discrete equation admitting several 
continuous limits.
In such a case,
the test must evidently be performed around every continuous limit.

\section{Conclusion
\label{sectionConclusion}}
\indent

We have tried to give some precise definitions and guidelines to make 
more systematic the search for discrete analogs of differential equations
integrable in the sense of Painlev\'e.

The discrete Painlev\'e test is now made of two main methods :
the singularity confinement method,
which seems intrinsically discrete,
and the present one,
which assumes an underlying continuous limit.
Both points of view (discrete first, continuous first) are useful since they
generate identical results on a sample of equations.

Rules of discretization for Lax pairs will, in future work,
be established and systematically applied to the numerous d--(Pn) or 
$q-$(Pn) candidates which still lack a discrete Lax pair.

Finally,
let us mention that the celebrated equation of class III of Chazy
\cite{ChazyThese}, 
$u''' - 2 u u'' + 3 u'^2=0$,
has recently been discretized~\cite{LabrunieConteChazyIII}
following the methods of this paper :
the no-log conditions are entirely obtained at order $n=2$,
and then they are all satisfied at least up to $n=16$.
However, this candidate d--(Chazy) has not yet been integrated.

{\it Acknowledgements}.
\indent

We thank A.~Magnus
for his references to the works of Shohat and Freud,
and J.-M.~Drouffe for the use of his efficient computer algebra language
AMP~\cite{Drouffe}.

Both authors acknowledge the financial support of the Tournesol grant
T 95/004.
M.~M.~acknowledges the financial support extended by 
Flanders's Federale Diensten voor Wetenschappelijke, Technische en Culturele
Aangelegenheden in the framework of the IUAP III no.~9.

\vfill \eject


\end{document}